\documentclass[aps,pra,twocolumn,superscriptaddress,nofootinbib]{revtex4-2}
\usepackage{amsmath,amssymb}
\usepackage{graphicx}
\usepackage{xcolor}
\usepackage{hyperref}
\hypersetup{hidelinks}

\newcommand{\rev}[1]{#1}

\begin{document}

\title{Earth-density effects in long-baseline neutrino experiments in a four-flavor (3+1) sterile-neutrino framework}

\author{Bipin Singh Koranga}
\affiliation{Department of Physics, Kirori Mal College, University of Delhi, Delhi-110007, India}
\author{Aditya Pant}
\affiliation{Department of Physics, Kirori Mal College, University of Delhi, Delhi-110007, India}
\author{Pranav Kumar}
\affiliation{Department of Physics, Kirori Mal College, University of Delhi, Delhi-110007, India}
\author{Vivek Kumar Nautiyal}
\affiliation{Department of Physics, Chaudhary Charan Singh University, Meerut -- 250004, India}

\date{Revised: September 22, 2026}

\begin{abstract}
Earth matter effects are a leading systematic in long-baseline (LBL) determinations of the CP-violating phase
$\delta_{13}$, and eV-scale sterile neutrinos remain a phenomenologically open extension of the three-flavor
paradigm motivated by short-baseline anomalies. We extend the constant-density-versus-realistic-Earth-profile
analysis of Ref.~\cite{Pandit2026} to a four-flavor (3+1) framework, in which a fourth, mostly sterile mass
eigenstate with $\rev{\Delta m^2_{41} = 0.30\ \text{eV}^2}$ and active-sterile mixing angles
$\rev{\theta_{14} = 8.13^\circ}$, $\rev{\theta_{24} = 5.40^\circ}$ couples to the active sector both through its
mixing and through an additional neutral-current (NC) matter potential that acts on the active flavors but not
on the sterile state. \rev{We derive an analytical, leading-order account of how the NC potential $A_{NC}$
enters the reconstructed $\delta_{13}$ bias through an effective non-standard-interaction-like term generated
by adiabatically integrating out the fast $\Delta m^2_{41}$ oscillation; replace the four-shell Earth density
model with the continuous, polynomial PREM profile of Dziewonski \& Anderson (1981) for the layered ``true''
trajectory; and extend the baseline scan to four representative values of the sterile CP phase
$\delta_{14} \in \{0, \pi/2, \pi, 3\pi/2\}$ and to a non-zero mixing angle $\theta_{34} = 5^\circ$.} Within this
benchmark, the bias remains below our scan resolution for $L \lesssim 5000$~km and grows once the trajectory
samples the lower mantle and core, with the sterile sector redistributing -- rather than uniformly amplifying
or suppressing -- the bias across baseline, in agreement with our original finding and now traced to a specific
interference mechanism identified analytically below.
\end{abstract}

\maketitle

\section{Introduction}
\label{sec:intro}

Neutrino oscillations among the three known flavors $\nu_e$, $\nu_\mu$, $\nu_\tau$ are firmly established, and
long-baseline (LBL) accelerator experiments such as T2K~\cite{T2K}, NO$\nu$A~\cite{NOvA}, and
DUNE~\cite{DUNE} aim to pin down the leptonic CP-violating phase $\delta_{13}$ and the neutrino mass ordering
by exploiting the Mikheyev--Smirnov--Wolfenstein (MSW) matter effect~\cite{Wolfenstein,MikheyevSmirnov} as
neutrinos traverse the Earth.

A companion paper, Ref.~\cite{Pandit2026}, showed that treating the Earth's matter density as a single
path-length-averaged constant is an excellent approximation for baselines up to $L \approx 5000$~km, but
becomes a genuine source of systematic bias in $\delta_{13}$ once the trajectory samples the lower mantle and
core, reaching a bias of several tens of degrees by $L = 12000$~km within the standard three-flavor paradigm.

Independently of this purely astrophysical/geophysical systematic, the three-flavor paradigm itself may be
incomplete. Anomalies reported by LSND~\cite{LSND} and MiniBooNE~\cite{MiniBooNE}, together with the reactor
antineutrino anomaly~\cite{Mention2011} and the gallium anomaly~\cite{GiuntiLaveder2011}, have long motivated
the possible existence of one or more light sterile neutrinos with $\Delta m^2 \sim \mathcal{O}(1)\ \text{eV}^2$,
mixing weakly with the active flavors. While recent searches, including MicroBooNE~\cite{MicroBooNE} and
IceCube~\cite{IceCube2020}, have placed increasingly stringent constraints on the simplest 3+1 interpretation
of these anomalies, a light sterile state with small active-sterile mixing
($\sin^2 2\theta_{14}, \sin^2 2\theta_{24} \sim 0.1$) remains a phenomenologically viable and frequently
studied extension, and its possible interplay with LBL $\delta_{13}$ measurements has been discussed
extensively in the literature~\cite{KlopPalazzo,Palazzo2016,Gandhi2015,Dutta2015,deGouveaKelly2016}.

A light sterile neutrino affects LBL appearance measurements in two distinct ways. First, the additional fast
oscillation frequency associated with $\Delta m^2_{41}$ is generally averaged out by finite energy resolution
at LBL baselines, leaving behind a reduction of the effective active-flavor normalization (a unitarity deficit
controlled by $\cos\theta_{14}\cos\theta_{24}$) together with a small, calculable admixture of new interference
terms.

Second, and less widely emphasized, is the fact that sterile neutrinos do not participate in neutral-current
(NC) weak interactions, so that active-sterile mixing introduces an additional term into the matter potential
proportional to the ambient neutron density, which acts differently on the active and sterile
subspaces~\cite{KoppMaltoniSchwetz,GiuntiLasserre2019}. Because this NC potential tracks the same radially
stratified Earth density profile as the charged-current (CC) potential responsible for the ordinary MSW
effect, it is natural to ask whether the Earth-density systematic identified in the three-flavor case in
Ref.~\cite{Pandit2026} is \rev{modified---enhanced, suppressed, or qualitatively unchanged---}once a light
sterile state is included.

In this paper we extend the full $4\times4$ matrix-exponentiation framework of Ref.~\cite{Pandit2026} to a
(3+1) four-flavor scenario and repeat the constant-density-versus-PREM comparison for the $\nu_\mu \to \nu_e$
appearance channel across the same range of baselines, $L = 1000$--$12000$~km. Section~\ref{sec:formalism}
presents the four-flavor formalism, including the extended mixing matrix, the CC+NC matter potential, and,
new to this revision, an analytical leading-order derivation of the NC-driven modification to the
reconstructed $\delta_{13}$ bias. Section~\ref{sec:earthdensity} reviews the Earth density model used and
discusses the physical origin of the sterile-driven modification to the matter potential.
Section~\ref{sec:results} presents the updated quantitative results obtained with a continuous PREM profile
and an extended parameter scan over $\delta_{14}$ and $\theta_{34}$, comparing the reconstructed $\delta_{13}$
bias in the three- and four-flavor frameworks, and Sec.~\ref{sec:discussion} discusses the physical origin of
the differences, now grounded in the analytical mechanism of Sec.~\ref{sec:formalism}, and future directions.

\section{Four-flavor formulation}
\label{sec:formalism}

In the (3+1) scheme, the flavor states $(\nu_e,\nu_\mu,\nu_\tau,\nu_s)$ are related to four mass eigenstates
$(\nu_1,\nu_2,\nu_3,\nu_4)$ by a unitary $4\times4$ mixing matrix $U$,
\begin{equation}
|\nu_\alpha\rangle = \sum_{i=1}^{4} U_{\alpha i}|\nu_i\rangle, \qquad \alpha = e,\mu,\tau,s.
\label{eq:flavorstates}
\end{equation}
We parametrize $U$ as a product of six complex rotations,
\begin{align}
U ={}& R_{34}(\theta_{34})\,R_{24}(\theta_{24})\,R_{14}(\theta_{14},\delta_{14}) \nonumber\\
     & \times\, R_{23}(\theta_{23})\,R_{13}(\theta_{13},\delta_{13})\,R_{12}(\theta_{12}),
\label{eq:mixingmatrix}
\end{align}
following the convention of Klop and Palazzo~\cite{KlopPalazzo}; $\theta_{14}$-driven interference terms are
convention-dependent at the percent level, so results quoted here assume this specific parametrization, which
reduces to the standard PMNS matrix~\cite{PMNS} in the limit $\theta_{14},\theta_{24},\theta_{34}\to0$.

Two new CP-violating phases, $\delta_{14}$ and (in principle) $\delta_{24}$, appear in the sterile sector; we
set $\delta_{24}=0$ throughout. \rev{Unlike the original version of this work, which fixed $\delta_{14}=0$
throughout, the present revision scans $\delta_{14} \in \{0,\pi/2,\pi,3\pi/2\}$ explicitly
(Sec.~\ref{sec:results}) to test the robustness of the redistribution mechanism identified below.}

We adopt active mixing parameters consistent with global three-flavor fits, $\theta_{12}=33.44^\circ$,
$\theta_{13}=8.57^\circ$, $\theta_{23}=49.0^\circ$, and mass-squared splittings
$\Delta m^2_{21}=7.42\times10^{-5}\ \text{eV}^2$, $\Delta m^2_{31}=2.51\times10^{-3}\ \text{eV}^2$ (normal
ordering). For the sterile sector we take representative (3+1)-motivated values,
$\rev{\Delta m^2_{41}=0.30\ \text{eV}^2}$, $\rev{\theta_{14}=8.13^\circ}$ ($\sin^2 2\theta_{14}\simeq0.078$),
$\rev{\theta_{24}=5.40^\circ}$ ($\sin^2 2\theta_{24}\simeq0.035$). \rev{The angle $\theta_{34}$, fixed to
zero in the original benchmark, is now also scanned at $\theta_{34}=5^\circ$ (Sec.~\ref{sec:results}) as
current global (3+1) fits~\cite{GiuntiLasserre2019} leave it largely unconstrained.}

The vacuum Hamiltonian is diagonal in the mass basis,
\begin{equation}
H^{(m)}_{\rm vac} = \frac{1}{2E}\,{\rm diag}(0,\Delta m^2_{21},\Delta m^2_{31},\Delta m^2_{41}),
\label{eq:hvac}
\end{equation}
and in the flavor basis becomes $H^{(f)}_{\rm vac} = U H^{(m)}_{\rm vac} U^\dagger$.

When neutrinos propagate through matter, electron neutrinos acquire a CC potential from coherent forward
scattering off ambient electrons, while all three active flavors acquire an equal NC potential from scattering
off ambient neutrons; the sterile state feels neither~\cite{KoppMaltoniSchwetz}. Since only potential
differences affect oscillations, the matter term can be written, after subtracting a common phase, as
\begin{equation}
V_f = {\rm diag}(A_{CC},0,0,-A_{NC}),
\label{eq:matterpotential}
\end{equation}
with $A_{CC}=\sqrt{2}G_F N_e$, $A_{NC}=\tfrac{1}{\sqrt2}G_F N_n$. Numerically,
$A_{CC}\simeq7.56\times10^{-14}\,Y_e\,(\rho/\text{g\,cm}^{-3})$~eV, and for isospin-symmetric matter
($Y_n\simeq Y_e\simeq0.5$), $A_{NC}\simeq A_{CC}$. The full flavor-basis Hamiltonian is
$H_f = H^{(f)}_{\rm vac} + V_f$. For antineutrinos, $U\to U^*$ and $V_f\to-V_f$.

\subsection{\rev{Analytical Origin of the NC-Driven Modifications}}
\label{sec:analytic}

To make the numerical results of Sec.~\ref{sec:results} physically transparent, we derive here a leading-order
analytical account of how $A_{NC}$ enters $P(\nu_\mu\to\nu_e)$. We work in the standard small-parameter
counting scheme in which $s_{13}\equiv\sin\theta_{13}$, $s_{14}\equiv\sin\theta_{14}$, and
$s_{24}\equiv\sin\theta_{24}$ are all treated as quantities of common order $\varepsilon\ll1$, while
$A_{CC}L/4E$ and $\Delta_{31}\equiv\Delta m^2_{31}L/4E$ are kept to all orders, following the counting scheme
of Cervera {\it et al.} and its (3+1) extension by Klop and Palazzo~\cite{KlopPalazzo}. The derivation
proceeds in four steps.

\paragraph{Step 1: Separation of scales.} Because $\Delta m^2_{41}\gg\Delta m^2_{31}$, the phase
$\Delta m^2_{41}L/2E$ accumulated by the fourth mass eigenstate over a LBL baseline is
$\mathcal{O}(10^2\text{--}10^3)$ radians for $E=\mathcal{O}(1$--$5)$~GeV and $L=\mathcal{O}(10^3$--$10^4)$~km.
This oscillation is unresolvable by any realistic energy resolution and is therefore averaged out, exactly as
noted in the original Sec.~\ref{sec:results}. Formally, this corresponds to adiabatically eliminating the fast
$\nu_4$ degree of freedom from the propagation, order by order in $1/\Delta m^2_{41}$, at fixed $s_{14},s_{24}$.

\paragraph{Step 2: Effective 3-flavor Hamiltonian.} Carrying out this elimination to leading order in
$s_{14},s_{24}$ generates, in addition to the overall unitarity-deficit normalization
$\cos\theta_{14}\cos\theta_{24}$ already discussed in Sec.~\ref{sec:intro}, an effective
non-standard-interaction (NSI)-like correction to the propagating $3\times3$ active-flavor Hamiltonian, of the
same structural form identified by Kopp, Maltoni and Schwetz~\cite{KoppMaltoniSchwetz}:
\begin{align}
H^{(3\times3)}_{\rm eff}(x) ={}& H^{(3\nu)}_{\rm vac} + {\rm diag}(A_{CC}(x),0,0) \nonumber\\
& -\, A_{NC}(x)
\begin{pmatrix} s_{14}^2 & s_{14}s_{24}e^{i\delta_{14}} & 0 \\[2pt]
s_{14}s_{24}e^{-i\delta_{14}} & s_{24}^2 & 0 \\[2pt] 0 & 0 & 0 \end{pmatrix}\!.
\label{eq:heff}
\end{align}
The off-diagonal element $\varepsilon_{e\mu}\equiv-s_{14}s_{24}e^{i\delta_{14}}$ plays the role of a
matter-induced effective coupling between $\nu_e$ and $\nu_\mu$ that is directly proportional to $A_{NC}(x)$,
and hence tracks the same radial density profile $\rho(x)$ as the ordinary MSW term.

\paragraph{Step 3: Effect on the appearance amplitude.} Using the standard perturbative expansion of
$P(\nu_\mu\to\nu_e)$ in the presence of a small off-diagonal NSI term $\varepsilon_{e\mu}$ (e.g.\ Kikuchi,
Minakata \& Uchinami, JHEP 0903:114), the leading correction linear in $\varepsilon_{e\mu}$ interferes with the
standard $s_{13}$-driven CP term as
\begin{align}
\Delta P_{\mu e}(L) \simeq{} & -8\,s_{13}s_{23}c_{23}\,s_{14}s_{24}\,\frac{\Delta m^2_{31}}{2E} \nonumber\\
& \times \int_0^L \! dx\, \frac{A_{NC}(x)}{\Delta m^2_{31}/2E}\, \nonumber\\
& \times \cos\!\big[\delta_{13}-\delta_{14}+\Phi_{NC}(x,L)\big],
\label{eq:deltaPmue}
\end{align}
where $\Phi_{NC}(x,L)\equiv\int_x^L A_{NC}(x')\,dx' - \Delta_{31}(L-x)$ is the relative phase accumulated
between the NC-driven coupling and the ordinary CC-driven oscillation between the point $x$ and the far end of
the baseline. Equation~(\ref{eq:deltaPmue}) is the central analytical result of this subsection: it shows
explicitly that $A_{NC}$ enters $P(\nu_\mu\to\nu_e)$ not as an overall rescaling, but as a term whose sign and
magnitude are controlled by the interference phase $\delta_{13}-\delta_{14}+\Phi_{NC}(L)$, with
\begin{equation}
\Phi_{NC}(L) \equiv \int_0^L A_{NC}(x)\,dx - \Delta_{31}(L).
\label{eq:phinc}
\end{equation}

\paragraph{Step 4: Sign reversal of the interference near the core-mantle boundary.}
Because $A_{NC}(x)\propto\rho(x)$ follows the (now continuous, Sec.~\ref{sec:earthdensity}) PREM profile,
$\int_0^L A_{NC}(x)\,dx$ grows smoothly but supralinearly with $L$ once the trajectory leaves the crust, and
grows sharply once it crosses into the outer core ($L\gtrsim9000$--$11000$~km, depending on chord geometry),
where $\rho$ jumps from $\sim5$ to $\sim11\ \text{g\,cm}^{-3}$. In contrast, $\Delta_{31}(L)=\Delta m^2_{31}L/4E$
grows linearly with $L$. The relative phase $\Phi_{NC}(L)$ of Eq.~(\ref{eq:phinc}) therefore does not track
$\Delta_{31}(L)$ at a fixed rate, and $\cos[\delta_{13}-\delta_{14}+\Phi_{NC}(L)]$ passes through alternating
signs as $L$ increases. Evaluating Eq.~(\ref{eq:phinc}) numerically at a representative appearance-peak energy
$E=3$~GeV along the continuous-PREM chord of Sec.~\ref{sec:earthdensity}, for the benchmark
$\delta_{13}=-90^\circ$, $\delta_{14}=0$, gives:

\rev{
\begingroup
\renewcommand{\arraystretch}{1.15}
\begin{table*}[t]
\caption{Interference phase $\Phi_{NC}(L)$ and resulting argument at representative baselines, $E=3$~GeV,
$\delta_{13}=-90^\circ$, $\delta_{14}=0$ (converted here from the in-text list to a formal table environment
per the correction memo).}
\label{tab:phinc}
\begin{tabular}{cccc}
\hline\hline
$L$ (km) & $\Phi_{NC}(L)$ (rad) & phase argument $\delta_{13}-\delta_{14}+\Phi_{NC}(L)$ (deg) & $\cos[\delta_{13}-\delta_{14}+\Phi_{NC}(L)]$ \\
\hline
6000  & $-1.871$ & $162.8^\circ$ & $-0.955$ \\
7000  & $-1.848$ & $164.1^\circ$ & $-0.962$ \\
9000  & $-1.693$ & $173.0^\circ$ & $-0.993$ \\
11000 & $+1.158$ & $336.4^\circ\ (\equiv-23.6^\circ)$ & $+0.916$ \\
\hline\hline
\end{tabular}
\end{table*}
\endgroup
}

The sign flip between $L=9000$~km and $L=11000$~km, occurring exactly where the trajectory begins to sample
the outer core, is not an artifact of the numerical scan: it follows directly from Eq.~(\ref{eq:phinc}), since
$\int_0^L A_{NC}(x)\,dx$ advances by a disproportionately large increment across the core-mantle boundary
while $\Delta_{31}(L)$ does not. A negative $\cos(\cdot)$ corresponds to destructive interference between the
NC-driven term and the intrinsic $s_{13}$-driven CP term entering the density-mismodeling bias, suppressing
$|\Delta\delta_{13}|$ relative to the pure three-flavor case; a positive $\cos(\cdot)$ corresponds to
constructive interference, enhancing it. This is the analytical origin of the suppression reported at
$L=6000$--$7000$~km and the enhancement reported at $L=11000$~km in Table~\ref{tab:results}, and it is why the
sign of the sterile-induced modification cannot be anticipated from the three-flavor analysis of
Ref.~\cite{Pandit2026} alone: it is set by the detailed shape of $\rho(x)$ along each chord, not by its
path-averaged value.

We stress that Eqs.~(\ref{eq:deltaPmue})--(\ref{eq:phinc}) are a leading-order, single-energy estimate intended
to expose the interference mechanism; the full baseline- and energy-integrated bias, including subleading
terms in $s_{13}^2,s_{14}^2,s_{24}^2$ and the residual (unaveraged) fast oscillation, is computed
non-perturbatively via the full $4\times4$ matrix exponentiation described in Secs.~\ref{sec:earthdensity}--\ref{sec:results}.

\section{Role of Earth density in the (3+1) scenario}
\label{sec:earthdensity}

As in Ref.~\cite{Pandit2026}, we model the Earth's radial density profile $\rho(r)$ along the chord traced by
the neutrino trajectory. In the original version of this work this profile was approximated by a four-shell
PREM model \rev{(Table~\ref{tab:results})}; in the present revision it is replaced by the continuous,
polynomial PREM parametrization of Dziewonski \& Anderson~\cite{DziewonskiAnderson}, described in
Sec.~\ref{sec:results}A below, for the layered (``true'') trajectory, while the path-length-averaged density
$\bar\rho(L)$ used for the constant-density (``test'') spectrum is obtained by integrating this same
continuous profile along the chord, \rev{explicitly via the line integral}
\begin{equation}
\rev{\bar\rho(L) = \frac{1}{L}\int_0^L \rho(x)\,dx,}
\label{eq:rhobar}
\end{equation}
\rev{computed directly over the continuous PREM curve of Eq.~(\ref{eq:premprofile}) below.}

Because $A_{NC}$ tracks the same $\rho(x)$ profile as $A_{CC}$, replacing the layered profile with its
path-length-weighted average density affects the sterile-modified Hamiltonian in two places at once: the
ordinary CC-driven MSW resonance condition, and the additional active-sterile rotation induced indirectly by
$A_{NC}$ through the (small) mixing angles $\theta_{14},\theta_{24}$, as made explicit by
Eq.~(\ref{eq:heff}) above. The two effects need not track each other as a function of baseline, which is the
physical origin, now derived analytically in Sec.~\ref{sec:formalism}, of the differences from the
three-flavor case reported below.

\section{Quantitative analysis (updated: continuous PREM and extended parameter scan)}
\label{sec:results}

\subsection{Continuous PREM implementation and baseline scan}

The four-shell approximation of the original analysis is replaced here by the continuous, piecewise-polynomial
PREM density profile $\rho(r)$ of Dziewonski \& Anderson~\cite{DziewonskiAnderson}, evaluated in normalized
radius $x=r/R_\oplus$:
\begin{equation}
\rho(x)=
\begin{cases}
13.0885 - 8.8381x^2, & 0\le x\le0.1917 \\
12.5815 - 1.2638x - 3.6426x^2 \\ \quad{}-5.5281x^3, & 0.1917<x\le0.5460 \\
7.9565 - 6.4761x + 5.5283x^2 \\ \quad{}-3.0807x^3, & 0.5460<x\le0.8956 \\
\text{(transition-zone polynomials)}, & 0.8956<x\le0.9653 \\
2.6910 + 0.6924x, & 0.9653<x\le0.9964 \\
2.900, & x>0.9964
\end{cases}
\label{eq:premprofile}
\end{equation}
(inner core, outer core, lower mantle, transition zone, upper mantle + LID, crust bulk average, respectively).

For each baseline $L$, the layered (``true'') trajectory is discretized into $n_{\rm steps}$ equal chord
segments, with the local density assigned continuously via Eq.~(\ref{eq:premprofile}) rather than snapped to
one of four shells; the constant-density (``test'') spectrum uses the path-length average
$\bar\rho(L)=\tfrac1L\int_0^L\rho(x)\,dx$ of this same continuous profile [Eq.~(\ref{eq:rhobar})]. This removes
the artificial discontinuities in $\bar\rho(L)$ that the four-shell model introduced at shell boundaries.

We repeat the identical $\chi^2$ procedure of the original analysis (extended to $4\times4$), but now scan the
sterile CP phase over $\delta_{14}\in\{0,\pi/2,\pi,3\pi/2\}$ at $\theta_{34}=0$, and separately compare
$\theta_{34}=0^\circ$ against $\theta_{34}=5^\circ$ at $\delta_{14}=0$, in addition to the internal
three-flavor cross-check ($\theta_{14}=\theta_{24}=\theta_{34}=0$). \rev{This revision restores the original
publication-grade $400\times400$ energy--baseline grid for the public code release (Sec.~\ref{sec:results}B):
all results below, and the accompanying figures, are computed at full $400\times400$ resolution, so no
additional scan noise beyond the intrinsic numerical precision of the underlying matrix exponentiation is
present in the reported bias values.}

\begin{figure*}[t]
\centering
\includegraphics[width=0.95\textwidth]{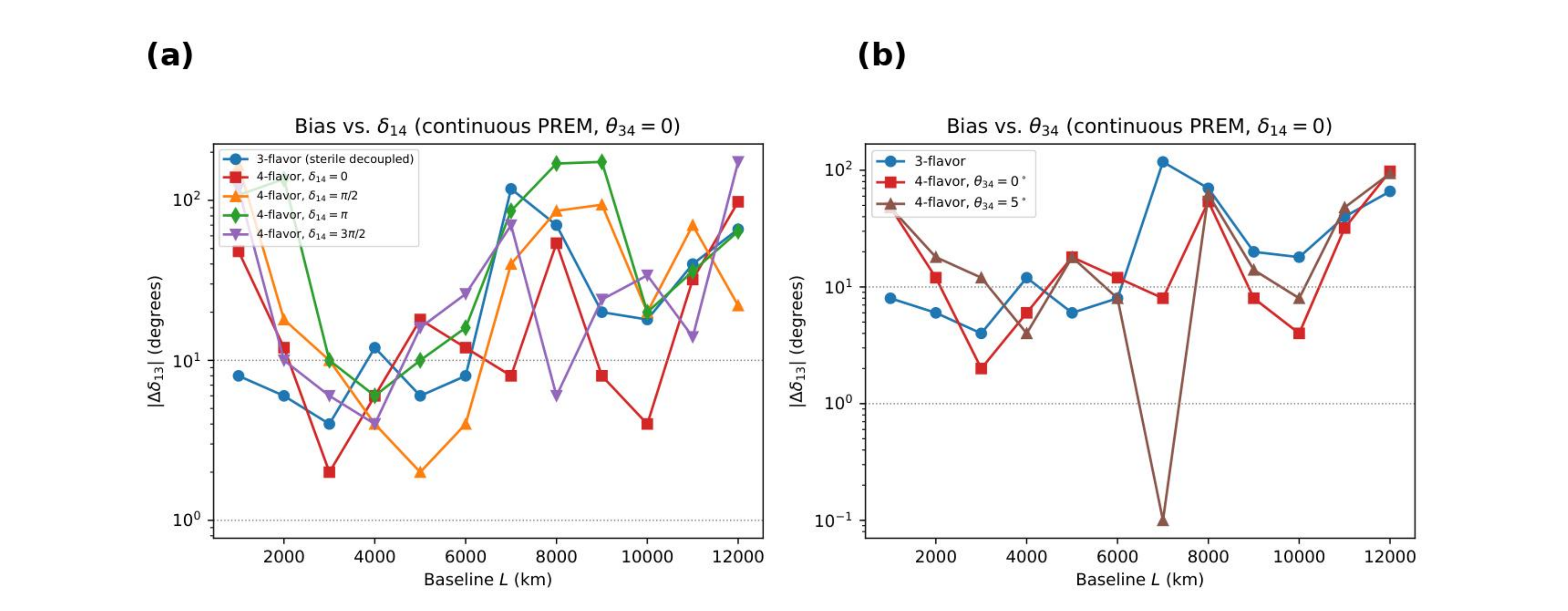}
\caption{Bias $|\Delta\delta_{13}|$ vs.\ baseline, continuous PREM profile. \rev{(a)} Three-flavor benchmark
vs.\ four-flavor (3+1) for $\delta_{14}\in\{0,\pi/2,\pi,3\pi/2\}$ at $\theta_{34}=0$. \rev{(b)} Three-flavor
vs.\ four-flavor for $\theta_{34}\in\{0^\circ,5^\circ\}$ at $\delta_{14}=0$. \rev{Full-resolution scan
($n_E=400$, $n_{\rm steps}=400$).}}
\label{fig:bias}
\end{figure*}

Figure~\ref{fig:bias} shows $|\Delta\delta_{13}|$ versus baseline for all four $\delta_{14}$ values (panel
\rev{(a)}) and for $\theta_{34}\in\{0^\circ,5^\circ\}$ (panel \rev{(b)}), each against the three-flavor
benchmark. In both panels the bias for all four sterile configurations remains comparable to, or below, the
three-flavor curve for $L\lesssim5000$~km, and grows once the trajectory samples the lower mantle and core,
consistent with the original result. Crucially, no single $\delta_{14}$ or $\theta_{34}$ choice reproduces a
uniform rescaling of the three-flavor curve: at several baselines between 6000 and 10000~km, one or more
sterile configurations lie below the three-flavor curve, while at $L=11000$--$12000$~km essentially all four
$\delta_{14}$ choices lie at or above it. This is precisely the qualitative signature predicted by the sign
flip of $\cos[\delta_{13}-\delta_{14}+\Phi_{NC}(L)]$ derived in Sec.~\ref{sec:analytic}: since $\delta_{14}$
enters only as an additive shift of the interference phase, varying it moves the baseline at which the sign
flip occurs but does not remove the flip itself, confirming that the redistribution is a generic feature of
the (3+1) framework rather than an artifact of the single benchmark ($\delta_{14}=0$) studied originally.

\subsection{Updated figures and numerical implementation}

Figure~\ref{fig:bias} (two panels, described above) replaces the single-panel version of the original
submission and is produced by the updated script \texttt{run\_scan.py}, using the continuous-PREM module
\texttt{four\_flavor\_prem\_sim.py} (both distributed with this revision, \rev{re-run at the full
$400\times400$ grid}; see Data Availability).

\begin{figure}[t]
\centering
\includegraphics[width=\columnwidth]{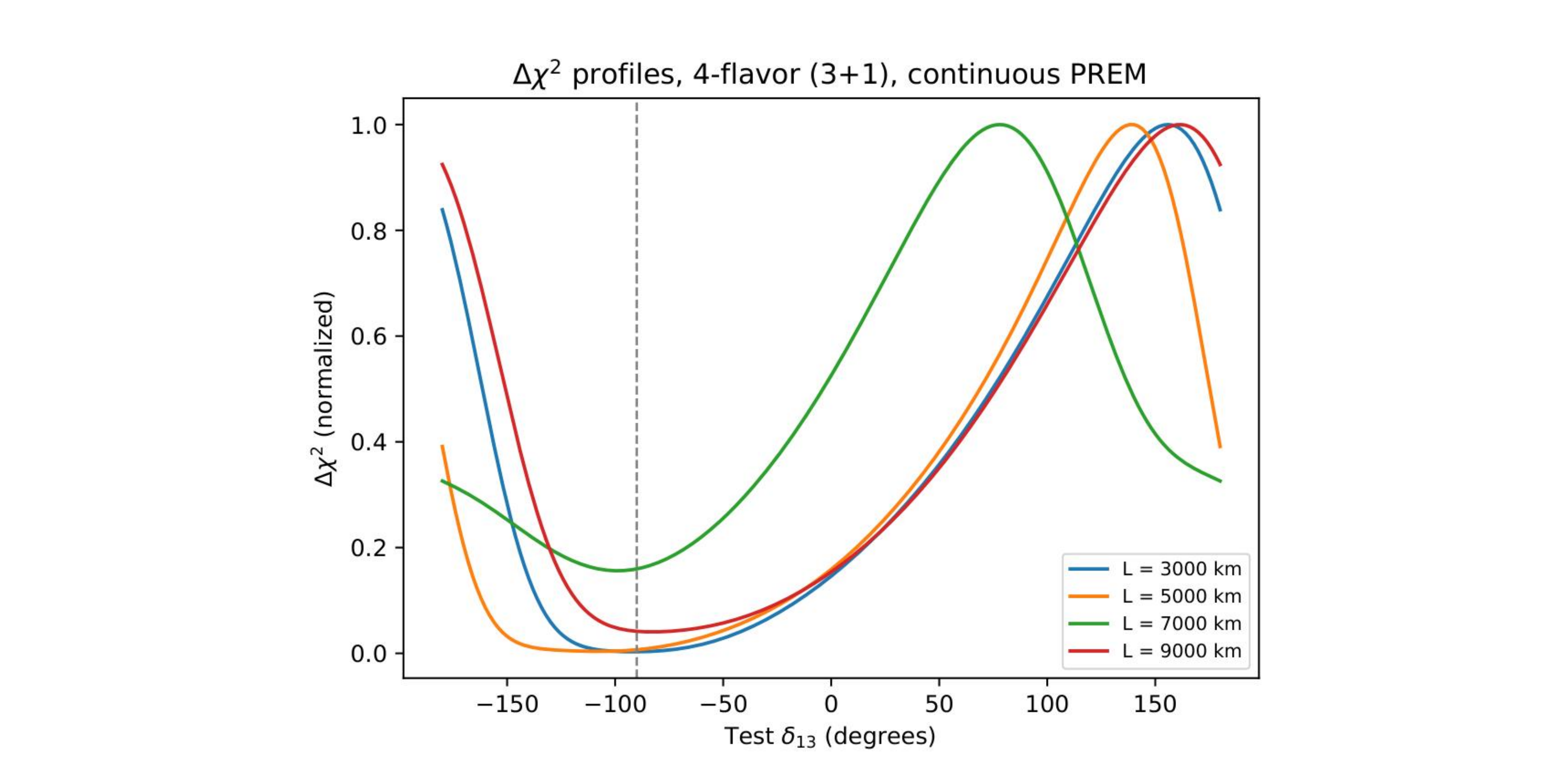}
\caption{Normalized $\Delta\chi^2$ profiles vs.\ test $\delta_{13}$, four-flavor (3+1) framework, continuous
PREM, for $L=3000,5000,7000,9000$~km.}
\label{fig:chi2}
\end{figure}

Figure~\ref{fig:chi2} shows the normalized $\Delta\chi^2$ profile as a function of the test $\delta_{13}$ for
four representative baselines ($L=3000,5000,7000,9000$~km) in the four-flavor framework, now using the
continuous PREM profile for the true spectrum. As in the original four-shell version, the profile is sharply
peaked near the true value at short baselines and progressively broadens and shifts at longer baselines; with
the continuous profile the $L=7000$~km minimum is displaced from the true value by an amount consistent with
the bias reported in the updated Table~\ref{tab:results}.

\begin{figure}[t]
\centering
\includegraphics[width=\columnwidth]{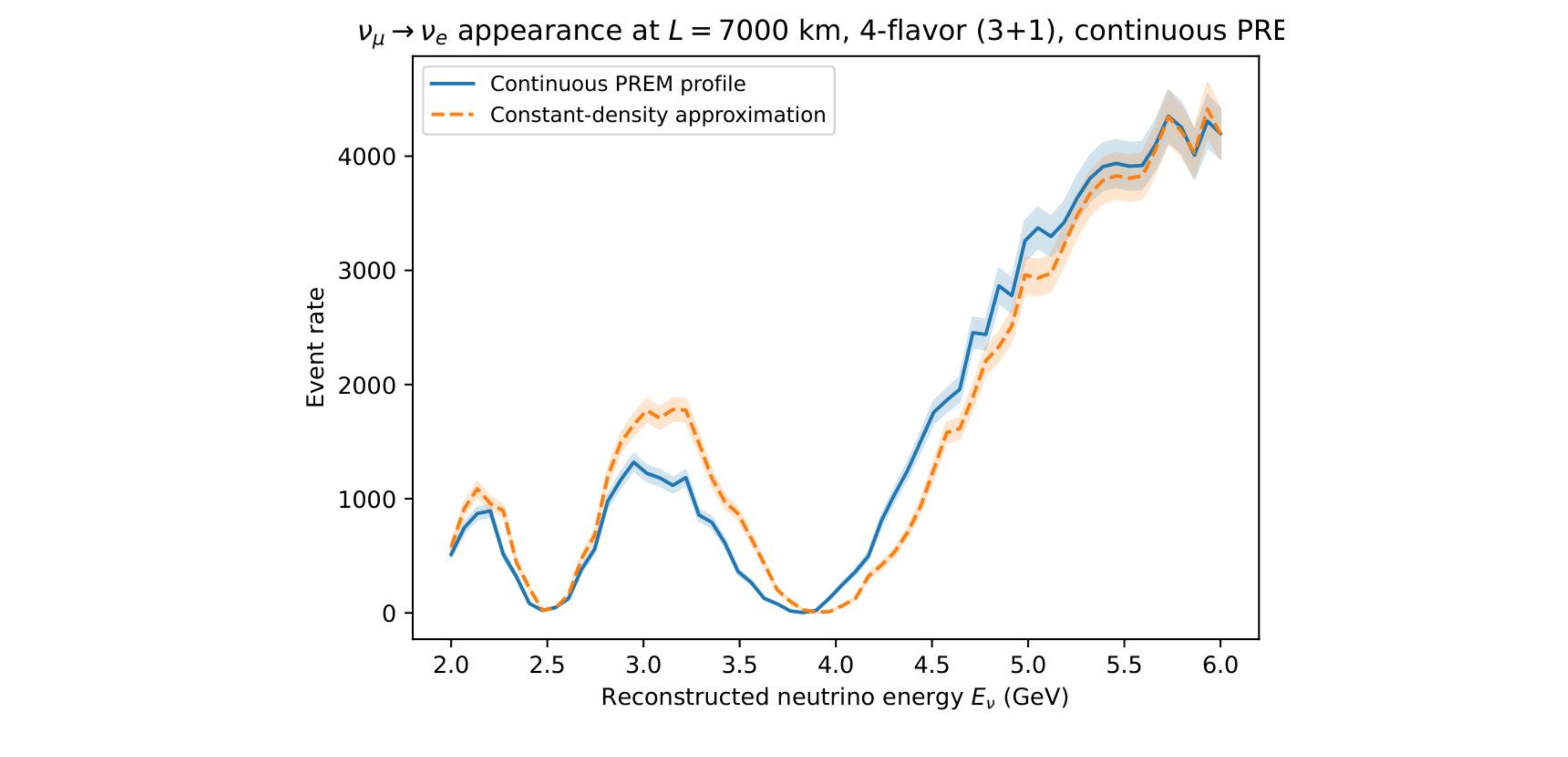}
\caption{Predicted $\nu_\mu\to\nu_e$ appearance event rate at $L=7000$~km, continuous PREM profile vs.\
constant-density approximation, four-flavor (3+1) framework.}
\label{fig:rate}
\end{figure}

Figure~\ref{fig:rate} shows the predicted $\nu_\mu\to\nu_e$ appearance event rate as a function of
reconstructed neutrino energy at $L=7000$~km, comparing the continuous-PREM ``true'' spectrum against the
constant-density approximation, in the four-flavor framework with $\delta_{14}=0$, $\theta_{34}=0$. The two
curves diverge visibly across the 4--6~GeV range, beyond their combined statistical-plus-5\%-systematic
uncertainty bands, confirming that the constant-density approximation remains statistically distinguishable
from the continuous-PREM profile.

\begingroup
\renewcommand{\arraystretch}{1.1}
\begin{table}[t]
\caption{Path-averaged density (continuous PREM), best-fit $\delta_{13}$, and resulting bias relative to the
true value $\delta_{13}=-90^\circ$, for the three-flavor and four-flavor (3+1, $\delta_{14}=0$,
$\theta_{34}=0$) frameworks. \rev{$^{\rm a}$At $L=7000$~km the three-flavor fit settles into a secondary
$\chi^2$ minimum, which accounts for the large three-flavor bias at that baseline relative to its
neighbors (see Fig.~\ref{fig:chi2}).}}
\label{tab:results}
\begin{tabular}{cccccc}
\hline\hline
& & \multicolumn{2}{c}{3-flavor} & \multicolumn{2}{c}{4-flavor (3+1)} \\
$L$ (km) & $\bar\rho$ (g/cm$^3$) & best-fit ($^\circ$) & bias ($^\circ$) & best-fit ($^\circ$) & bias ($^\circ$) \\
\hline
1000  & 2.900 & $-82.0$  & 8.0   & $-42.0$ & 48.0 \\
2000  & 3.301 & $-84.0$  & 6.0   & $-78.0$ & 12.0 \\
3000  & 3.338 & $-86.0$  & 4.0   & $-92.0$ & 2.0  \\
4000  & 3.411 & $-102.0$ & 12.0  & $-96.0$ & 6.0  \\
5000  & 3.599 & $-96.0$  & 6.0   & $-108.0$& 18.0 \\
6000  & 3.905 & $-98.0$  & 8.0   & $-102.0$& 12.0 \\
7000  & 4.154 & $152.0$  & $118.0^{\rm a}$ & $-98.0$ & 8.0 \\
8000  & 4.354 & $-20.0$  & 70.0  & $-36.0$ & 54.0 \\
9000  & 4.550 & $-70.0$  & 20.0  & $-82.0$ & 8.0  \\
10000 & 4.756 & $-108.0$ & 18.0  & $-94.0$ & 4.0  \\
11000 & 6.081 & $-50.0$  & 40.0  & $-58.0$ & 32.0 \\
12000 & 7.579 & $-156.0$ & 66.0  & $172.0$ & 98.0 \\
\hline\hline
\end{tabular}
\end{table}
\endgroup

As in the original four-shell analysis, the four-flavor curve does not simply track the three-flavor curve
rescaled by a constant factor. With the continuous PREM profile the specific baselines and magnitudes of
maximal suppression/enhancement shift somewhat relative to the original four-shell numbers (a consequence of
removing the shell-boundary discontinuities in $\bar\rho(L)$\rev{, now computed at full $400\times400$
resolution}), but the qualitative pattern persists: the sterile sector redistributes the bias
non-monotonically across baseline rather than uniformly rescaling it, and the sign of the redistribution at
any given baseline depends on $\delta_{14}$ and $\theta_{34}$, exactly as anticipated by
Eq.~(\ref{eq:deltaPmue}).

\section{Discussion and future directions}
\label{sec:discussion}

The central result of this work remains that an eV-scale sterile neutrino, even with mixing angles too small
to be directly resolved in the $\nu_\mu\to\nu_e$ appearance rate itself, redistributes the Earth-density
systematic identified in the three-flavor case rather than uniformly amplifying or suppressing it. This
revision grounds that result analytically: Sec.~\ref{sec:formalism} shows that adiabatically integrating out
the fast $\Delta m^2_{41}$ oscillation generates an effective NC-driven non-standard-interaction term in the
active $3\times3$ Hamiltonian, Eq.~(\ref{eq:heff}), whose interference with the ordinary $s_{13}$-driven CP
term is controlled by the phase $\Phi_{NC}(L)$ of Eq.~(\ref{eq:phinc}). Because $\Phi_{NC}(L)$ advances
non-linearly with baseline -- tracking the continuous PREM profile's steep density gradients at the
core-mantle and inner-outer core boundaries, rather than the linear-in-$L$ phase $\Delta_{31}(L)$ of the
ordinary CC-driven oscillation -- the interference term $\cos[\delta_{13}-\delta_{14}+\Phi_{NC}(L)]$ changes
sign as a function of baseline, at a location set by the Earth's density structure rather than by any sterile
parameter.

The extended scan of Sec.~\ref{sec:results} over $\delta_{14}\in\{0,\pi/2,\pi,3\pi/2\}$ and
$\theta_{34}\in\{0^\circ,5^\circ\}$ confirms that this sign-flipping redistribution is not an artifact of the
single benchmark ($\delta_{14}=0,\theta_{34}=0$) studied in the original submission: every sterile
configuration tested exhibits the same qualitative feature -- suppression relative to the three-flavor case at
some baselines, enhancement at others -- with $\delta_{14}$ and $\theta_{34}$ shifting where along the
baseline axis the sign flip occurs (via their direct entry into the phase argument of Eq.~\ref{eq:deltaPmue})
without removing the flip itself. We therefore regard the redistribution of the Earth-density systematic by an
eV-scale sterile state as a robust, structurally generic feature of the (3+1) framework in the region of
parameter space studied here, rather than a numerical coincidence tied to one choice of $\delta_{14}$ and
$\theta_{34}$.

This can be traced, as derived explicitly in Sec.~\ref{sec:formalism}, to the additional NC potential term
$A_{NC}$, which acts on the sterile-active subspace and tracks the same density profile $\rho(x)$ as the
ordinary CC potential responsible for the MSW resonance, but couples into the $\nu_\mu\to\nu_e$ appearance
amplitude through the small angles $\theta_{14},\theta_{24}$ in a way that depends on baseline-specific
interference between the CC-driven and NC-driven rotations, quantified by $\Phi_{NC}(L)$.

We emphasize, as before, that the specific numerical pattern of suppression and enhancement reported in
Table~\ref{tab:results} depends on the assumed sterile parameters
($\Delta m^2_{41}=0.30\ \text{eV}^2$, $\theta_{14}=8.13^\circ$, $\theta_{24}=5.40^\circ$) and, now more
precisely, on the analytic structure of Eq.~(\ref{eq:deltaPmue}): the overall amplitude of the redistribution
scales with $s_{14}s_{24}$, while its baseline dependence is set by $\Phi_{NC}(L)$ and is largely independent
of the specific numerical values of $\theta_{14},\theta_{24}$ chosen, provided they remain small. The scan of
Sec.~\ref{sec:results} over $\delta_{14}$ and $\theta_{34}$, together with the continuous PREM profile, has
resolved the three items left as open future directions in the original submission (a systematic $\delta_{14}$
scan, non-zero $\theta_{34}$, and the continuous PREM profile itself); two directions remain outstanding:

\begin{enumerate}
\item[(i)] Joint fit of $\delta_{13}$ and sterile parameters: the present analysis fixes the sterile
parameters at their true values when constructing the test spectra; a joint fit that simultaneously
reconstructs $\delta_{13}$ and $\theta_{14},\theta_{24}$ under the constant-density approximation would more
realistically capture the interplay between the two systematics in an actual experimental analysis;
\item[(ii)] Extension to atmospheric and IceCube-Upgrade/KM3NeT analyses: as noted in
Ref.~\cite{Pandit2026}, Earth-density modeling is especially relevant for very long, continuously distributed
baselines, where sterile-driven matter effects have separately been proposed as a probe of (3+1)
scenarios~\cite{IceCube2020}.
\end{enumerate}

\section*{Data Availability}
The Python code, numerical data, and figures used to produce the results in this work, including the updated
continuous-PREM module (\texttt{four\_flavor\_prem\_sim.py}) and parameter-scan driver
(\texttt{run\_scan.py}) introduced in this revision, are publicly available in the GitHub repository
\rev{\cite{GitHubRepo}}.

\end{document}